# CrystalGPU: Transparent and Efficient Utilization of GPU Power

Abdullah Gharaibeh, Samer Al-Kiswany, Matei Ripeanu
*Electrical and Computer Engineering Department*
*The University of British Columbia*
*{abdullah, samera, matei}@ece.ubc.ca*

## Abstract

*General-purpose computing on graphics processing units (GPGPU) has recently gained considerable attention in various domains such as bioinformatics, databases and distributed computing. GPGPU is based on using the GPU as a co-processor accelerator to offload computationally-intensive tasks from the CPU. This study starts from the observation that a number of GPU features (such as overlapping communication and computation, short lived buffer reuse, and harnessing multi-GPU systems) can be abstracted and reused across different GPGPU applications.*

*This paper describes CrystalGPU, a modular framework that transparently enables applications to exploit a number of GPU optimizations. Our evaluation shows that CrystalGPU enables up to 16x speedup gains on synthetic benchmarks, while introducing negligible latency overhead.*

## 1 Introduction

Today's GPUs offer drastic reduction in computational costs compared with existing traditional CPUs. These savings encourage using GPUs as hardware accelerators to support computationally-intensive applications like, for example, scientific applications [1] or distributed computing middleware [2]. In particular, GPUs are well suited to support stream processing applications, that is applications that process a continuous stream of data units, where the processing of a single data unit may exhibit intrinsic parallelism that can be exploited by offloading the computation to the GPU. Examples of such applications include video encoding/decoding [3] and deep packet inspection [4, 7].

Recent GPU models include a number of new capabilities that not only enable even lower computational costs, but also make GPUs an interesting computational platform for fine-granularity, latency-sensitive applications. For example, new GPU models have the ability to overlap computation and the communication with the host; thus offering the opportunity to hide communication overheads, a significant source of overheads for stream processing applications. Additionally, the newly introduced dual-GPU devices (e.g., NVIDIA's GeForce 9800 GX2) offer an additional level of parallelism by placing two GPUs on the same graphics card, and thus offer an opportunity to harness additional computing capacity.

However, the efficient utilization of these capabilities is a challenging task that requires careful management of GPU resources by the application developer that leads to extra development effort and additional application complexity [5]. This is mainly due to the peculiar characteristics of current GPU architectures and development environments. For example, to effectively overlap computation and communication, the application needs to asynchronously launch a number of independent computation and data transfer tasks, which require the application to maintain additional state information to keep track of these asynchronous operations. Also, to employ dual-GPU devices, the application has to manually detect the available devices, and spread the work among them.

This project proposes CrystalGPU, an execution framework that aims to simplify the task of GPU application developers by providing a higher level of abstraction, while, concurrently, maximizing GPU utilization. In particular, the framework runs entirely on the host machine and provides a layer between the application and the GPU, managing the execution of GPU operations (e.g., transferring data from/to the GPU and starting computations on the GPU). To this end, the framework seeks to transparently enable three optimizations that can speedup GPU applications:

- First, reusing GPU memory buffers to amortize the cost of allocation and deallocation of short lived buffers. Al-Kiswany et al. [2] demonstrate that the percentage of total execution time spent on memory buffer allocations can be as high as 80% of the execution time.

- Second, new GPU architectures and programming environments allow applications to overlap the communication and computation phases. Al-Kiswany et al. [2] show that the communication overhead can be as high as 40% for data-intensive applications. Other studies have also demonstrated the significance of communication overhead when using the GPU [6-8]. In these cases, overlapping computation and communication creates the opportunity to hide the communication cost, maximizing the utilization of the GPU computational power as a result.
- Third, multi-GPU systems have become common: Dual-GPU boards are commercially available and research groups started to prototype GPU clusters that assemble even more GPUs on the same workstation creating, as a result, an opportunity, to aggregate additional low-cost computing capacity.

The contributions of this work are summarized as follows:
- First, we present a high-level abstraction that aims to transparently maximize the utilization of a number of GPU features, while simplifying application developers' task. We contend that, in addition to the processor/co-processor scenario we analyze in detail, the abstraction we propose can be easily applied in other situations that involve massively multicore hardware deployments, for example asymmetric multicores like IBM's Cell Broadband Engine Architecture.
- Second, our prototype of this abstraction brings valuable performance gains in a realistic usage scenario where the communication overhead is significant. Our prototype demonstrates an up to 16x performance improvements when integrated with StoreGPU: a GPU based library that accelerates a number of hashing based primitives commonly used in distributed storage systems.
- Third, we make CrystalGPU available to the community[1]. This framework can be used by a wide range of applications including stream-processing applications and embarrassingly-parallel data processing.

The rest of this paper is organized as follows. Section 2 presents related background to GPU programming and the type of applications we target. Section 3 discusses CrystalGPU's API. Section 4 presents the framework's design. Section 5 presents our experimental results. Section 6 discusses related work. We conclude in Section 7.

## 2 Background

In this section, we present background related to GPU programming model (Section 2.1), and describe the type of applications that CrystalGPU targets (Section 2.2).

### 2.1 GPU Programming

Recently, GPUs underwent a radical evolution from a fixed-function, special-purpose pipeline dedicated to graphics applications to a fully programmable SIMD processor [9]. These changes transformed the GPUs into powerful engines that support offloading computations beyond graphics operations.

In general, offloading computation to the GPU is a three-stage process. First, *transferring input data to the GPU's internal memory*. GPUs do not have direct access to the host's main memory; rather they can only access there onboard memory. Consequently, the application has to explicitly allocate buffers on the GPU local memory, and transfer the data to them through an I/O interface, such as the host's PCI-Express bus. Further, data transfer operations are performed using direct memory access (DMA) engine, which requires the application to have the data placed in the host's pinned (i.e., non-pageable) memory. Therefore, it is important for the application to use pinned buffers for the data from the beginning, otherwise the device driver will perform an additional copy to an internal pinned buffer before transferring the data to the GPU.

Second, *processing*. Once the data is transferred to the GPU's internal memory, the application starts the computation by invoking the corresponding 'kernel', a function that when called is executed on the GPU. A number of optimizations can be applied at the kernel level (e.g., to enable efficient utilization of the GPU's internal memory architecture, minimizing thread divergence, etc.), however, these application-level optimizations are beyond the scope of our work as we focus on higher-level optimizations agnostic to the details of the kernel implementation.

Third, *transferring the results from the GPU to the host's main memory*. After the kernel finishes execution, the application explicitly transfers the results through the I/O channel back to the host's main memory.

This three-stage process, collectively named herein a 'job', is repeated by a stream processing application for each new data block to be processed. Note that due to limitations in current GPGPU programming environments, a job can only be executed on a single GPU even in dual-GPU cards. Therefore, using multi-GPU architectures requires the application to explicitly divide the data into two data sets, and perform two transfers and kernel invocations, one for each GPU.

---
[1] CrystalGPU is an open source project, the code can be found at http://netsyslab.ece.ubc.ca

## 2.2 CrystalGPU Applications

In this section, we discuss the type of applications that CrystalGPU targets. We classify the candidate applications as stream- and batch-processing applications depending on their real-time constraints.

*A. Stream processing applications.* Streaming applications sequentially process a steady stream of small sized data units. Such applications pose real-time constraints that entail low-latency overheads. In this category, parallelism is based on two opportunities. First, the processing of each data unit represents a single job that may have some intrinsic parallelism. Second, accumulating a number of data units and processing them in parallel as a single job offers the tradeoff of increasing parallelism, and consequently the processing throughput, while also increasing the processing latency for individual jobs.

One example in this category is video encoding/decoding. Recently, high-definition television (HDTV) broadcast has become popular. Such technology enables higher resolution than traditional broadcast formats, however it requires efficient video compression/decompression mechanisms to reduce transmission costs over the network. Van der Laan et al., [3] for example, present a GPU-accelerated video codec library that implements a number of common video compression/decompression techniques such as block motion compensation and frame arithmetic.

Another example is using the GPU to accelerate distributed systems middleware-level techniques. For instance, a number of distributed storage systems (e.g., Venti [10], OceanStore [11]) employ a technique known as content-addressable storage: files are divided in chunks, which, in turn, are identified based on their content by using the hash of the chunk as the its identifier. Content-addressable storage simplifies the separation of file and chunk metadata, and facilitates the identification and elimination of duplicate chunks, hence minimizing the amount of data that storage systems need to manage and transfer. However, the overhead of computing chunk hashes may limit performance. Al-Kiswany et al. [2] presents StoreGPU, a library that accelerates a number of hashing based primitives that support content addressable storage.

Deep packet inspection (DPI) is yet another example of streaming applications. A practical DPI solution must support high-speed packet inspection and impose low latency overheads. For example, the GPU can be used to accelerate a number of computationally expensive DPI operations such as header classification and signature matching algorithms [4, 7].

In the above presented use cases, the GPU is used to sequentially carry computations on a stream of data blocks (frames in the first, data chunks in the second, and packets in the third use case). CrystalGPU reduce the latency of such stream-processing applications by hiding the time spent on buffer allocations and data transfers to/from the GPU. Further, it enables increased throughput by efficiently harnessing multi-GPU architectures.

*B. Batch processing applications.* We include in this category applications for which a large number of independent jobs are available for processing at any point in time and that individual jobs do not have latency constraints. In this case, parallelism can be easily extracted by bundling individual jobs into batches. Folding@Home [12] is an example of this category: it aims to analyze a large number of chunks of biological data, chunks are independent and are all available at the beginning of the analysis.

For this category, CrystalGPU can enhance the utilization of the GPU by overlapping the computation for one batch with the data transfer for the next one; further, it enables transparent utilization of multi-GPU systems where multiple batches can be processed in parallel.

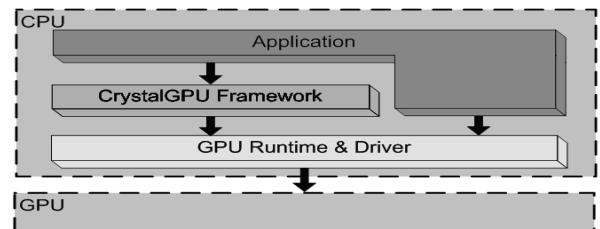

**Figure 1: CrystalGPU is a layer between the application and the GPU runtime.**

## 3 CrystalGPU API

CrystalGPU framework is a management layer between the application and the GPU runtime (Figure 1). The framework's API aims for generality to facilitate support for wide range of applications. We achieve this goal via an interface that is agnostic to the upper level application.

The framework's API is influenced by the GPU's programming model described in section 2.1. In general, job execution on the GPU is enabled by providing the application mechanisms to *(i)* define the input data, *(ii)* define the execution kernel, and *(iii)* claim the results. To facilitate these mechanisms CrystalGPU defines the **job** abstract data type (Figure 2) which specifies the main parameters needed to describe a job.

```
typedef struct job_s {
   void *h_input;
   void *h_output;
   void *d_input;
   void *d_output;
   int input_size;
   int output_size;
   ...
   void(*kernel_func)(struct job_s *);
   void(*callback_func)(struct job_s *);
} job_t;
```
**Figure 2: Job data type (C-Style).**

The *h_input* member of the **job** data type specifies the host buffer to be transferred to the destination GPU buffer *d_input*; likewise, *d_output* specifies the GPU results buffer to be transferred back to the host destination buffer *h_output*. Additionally, *kernel_func* specifies the application-specific kernel to be executed on the GPU, while *callback_func* specifies the callback function to be invoked when the job is done.

Figure 3 illustrates the framework's public interface. When initializing the framework, the application declares the maximum input/output buffer sizes it will use. The framework, accordingly, creates a buffer pool by pre-allocating a number of host and device buffers. The rationale behind the buffer pool is twofold. First, experience shows that GPU buffer allocation is expensive, especially for small scale computations [2]. By pre-allocating GPU buffers once at the beginning, the allocation time overhead gets amortized. In section 5, we show that this optimization provides a significant speedup. Second, the GPU driver always needs to use buffers in pinned memory for DMA transfers between the host and the GPU. As a result, pre-allocating buffers directly in pinned memory, and making them available for the application in advance, saves the GPU driver from an extra memory copy to an internal pinned memory buffer, thus reducing data transfer overhead.

The API enables the application to acquire and release pre-allocated buffers (encapsulated within a *job* instance) from the pool (*job_get*, *job_put*), submit a job to the framework to relay it to the GPU (*job_submit*), and to synchronously query or block on the status of a job (*job_query*, *job_synch*). Note that the later is in addition to the asynchronous notification enabled by the callback function described before; hence, allowing further flexibility for the application.

| Interface | Description |
|---|---|
| `init(max_input_size, max_output_size)` | Initializes the framework, the parameters define the maximum size of the input/output pre-allocated buffers. |
| `finalize()` | Cleans the framework's state. |
| `job_t * job_job_get()` | Gets a free job from the free pre-allocated pool. |
| `job_put(job_t *job)` | Returns back a job to the free pool. |
| `job_submit(job_t *job)` | Submits a job to be dispatched to the GPU. |
| `job_synch(job_t *job)` | Blocks till the designated job finishes execution. |
| `job_query(job_t *job)` | Returns true if the job has finished execution, false otherwise. |

**Figure 3: CrystalGPU public API.**

## 4 CrystalGPU Design

Crystal CPU design aims for efficiency to maximize the utilization of the GPU. The design makes minimal assumptions about the internal GPU architecture, while focusing on avoiding spurious data copies and concurrency bottlenecks.

We have implemented a prototype of CrystalGPU framework. The prototype is built atop of NVIDIA's CUDA toolkit [13]; however other alternative low-level GPU development toolkits (e.g., RapidMind [14] and OpenCL [15]) can also be used.

The rest of this section discusses the challenges that influenced CrystalGPU internal design (section 4.1) and the framework's internal details (section 4.2).

### 4.1 Design Challenges

CUDA supports asynchronous data transfers and kernel launches, which opens the opportunity to overlap data transfers and kernel execution across jobs. However, realizing the asynchronous operations' potential entails a number of challenges:

- *Lack of asynchronous notification mechanism.* Although CUDA supports asynchronous operations, it does not provide a notification mechanism to asynchronously interrupt the application when a GPU operation completes. Consequently, the application is required to periodically poll the CUDA run time library to query the status of a job.
- *Interleaved jobs execution.* CUDA enables asynchronous mode for data transfer operation and kernel executions. This asynchronous mode can be used to overlap the data transfer of one job with the kernel execution of another, increasing the system overall throughput accordingly. However, to effectively achieve this overlap, the application needs to explicitly interleave the execution of a number of jobs (at least three). Interleaving the execution of jobs is done by asynchronously issuing input data transfer operations for all available jobs, invoking the kernel asynchronously for all jobs, and finally issuing the asynchronous result transfer back from the GPU for all the jobs. This interleaved job execution dramatically complicates application design and development.

- *Primitive support for multi-GPU systems*. While recent GPU cards are fabricated with multiple GPU devices deployed on the same card (similarly multiple GPU cards can be deployed at one workstation), CUDA provides only primitive support for such multi-GPU systems. In order to use multiple GPUs, the application should decide, for each job, which GPU device it will be executed on. To this end, the application must track GPU device load, and implement a load balancing mechanism to achieve maximum hardware utilization.
- *Additional state to maintain*. Finally, given the asynchronous nature of GPU operations, the application needs to keep track of jobs execution status throughout the job lifetime.

The next section describes CrystalGPU design that transparently addresses these challenges.

## 4.2 CrystalGPU Internal Design

CrystalGPU design comprises a single driving module named the *master*. The master module employs a number of host-side master threads, each assigned to manage one GPU device.

The rationale behind this assignment is twofold. First, each master thread is responsible for querying its device for job completion status, and asynchronously notifying the application, using the callback function, once the job is done. This allows the client to make progress on the CPU in parallel; further, it relieves the application from job execution state bookkeeping. Second, having a dedicated control thread for each GPU facilitates transparent multi-GPU systems support. As detailed in the next paragraphs, the application submits a job to a shared outstanding queue, later the job is transparently handled by one of the master threads in such a way that balances the load across GPUs. We note that this multithreaded design requires a multi-core CPU to enable maximum parallelism across the master threads as well as the application's host-side threads.

The application requests services from the framework by submitting jobs and waiting for callbacks. The status of a job is maintained by the master using three queues. First, the *idle queue* maintains empty job instances (with pre-allocated buffers). Second, the *outstanding queue* contains ready-to-run jobs submitted by the application, but not dispatched to the GPUs yet. Third, *running queue* contains the jobs that are currently being executed on the GPUs.

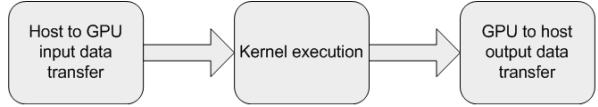

**Figure 4: CrystalGPU execution pipeline.**

A master thread is chosen in a round robin scheme to execute the next job in the outstanding queue. The execution of a job forms a three stage pipeline (Figure 4) that corresponds to the stages described in section 2. Therefore, having three or more jobs available always for execution is necessary to fill the pipeline and maximize the utilization of the GPU by overlapping the communication of one job with the computation of another. Note that the framework does not control the time spent on each stage of the pipeline as this is related to the characteristics of the application. As a result, having a balanced pipeline (i.e., one with equal communication and computation overheads) is necessary to efficiently exploit the opportunity of overlapping communication and computation.

## 5 Experimental Evaluation

We evaluate our prototype using a NVIDIA GeForce 9800GX2 GPU (released March 2008). The card is a dual GPU composed of two internal GeForce 8800 GTS GPUs. The host is Intel quad-core 2.66 GHz machine with PCI Express 2.0 x16 bus. Note that the two internal GPUs share the same PCI bus.

The experiments aim to evaluate the overhead introduced by the framework (section 5.1) as well as the performance gains enabled by the framework in one of the usage scenarios described earlier, namely when integrated with the StoreGPU library (section 5.2).

Finally, for all experiments, we report averages over at least 30 runs with 95% confidence intervals.

### 5.1 Overhead

The first experiment aims to evaluate the overhead introduced by CrystalGPU. To this end, we run an experiment that measures the time spent in the CrystalGPU layer while executing a single job with and without the framework. The job involves: *(i)* transferring a specific data size to the GPU, *(ii)* executing a dummy kernel with insignificant execution time, and (*iii*) transferring the same amount of data back to the CPU.

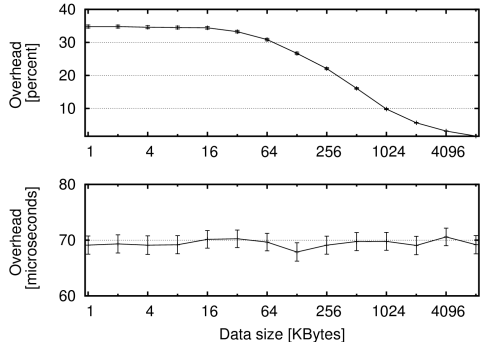

**Figure 5: CrystalGPU overhead while executing a single job with very low execution time. Top plot: percentage of total execution time. Bottom plot: absolute microseconds spent in CrystalGPU layer.**

Figure 5 shows the overhead while varying the data size. The results show that the framework introduces constant latency overhead irrespective of the data size (slightly less than 70μs), the result of threading and managing jobs in queues. This overhead corresponds to a significant share of the total execution time for small data sizes (up to 35%).

Nevertheless, in a realistic scenario, like the one we describe in the next section, a single job execution time is in the order of milliseconds, even for small data sizes. This is due to the larger kernel execution time in real applications. This renders the framework's latency overhead insignificant. Further, for the type of applications we target in which we have more than one job submitted sequentially back to back or as a group, the overhead introduced by the framework on one job is completely hidden by the kernel execution time of a former one as only one job can be executed on the GPU processing units at a time.

## 5.2 Application-level Performance

This section evaluates the performance gains achieved by StoreGPU library when integrated with CrystalGPU framework. The integration required modifying the StoreGPU library from directly calling CUDA functions to calling CrystalGPU's interface. These changes were fairly localized and minimal.

We note that the workload StoreGPU serves (e.g., computing the hash value for a stream of large data blocks) is similar to other stream processing applications like DPI and video encoding/decoding.

StoreGPU accelerates a number of hashing-based primitives popular in storage systems that use content-based addressing and is optimized to run on the GPU.

Here, however, we do not explore these optimizations; rather, we use a fully optimized StoreGPU version to explore the performance gains enabled by three techniques at the CrystalGPU framework level: (*i*) reusing memory buffers, (*ii*) overlapping computation and communication, and (*iii*) using multi-GPU architectures.

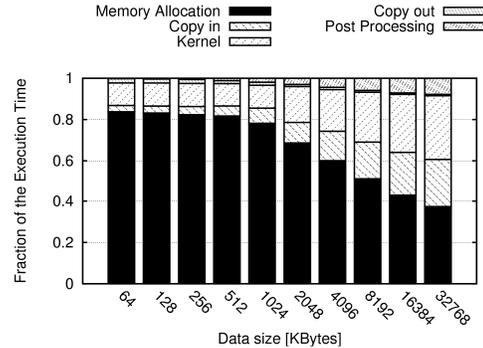

**Figure 6: Fraction of execution time spent on each stage for SHA1 direct hashing algorithm.**

Figure 6 demonstrates the percentage of execution time spent at each execution stage using the original implementation provided by the StoreGPU library (direct hashing based on SHA1). The hash computation is carried in five main stages: GPU and CPU pinned memory allocation, data copy into the GPU, kernel execution, results copy out to the host machine, and post processing on the CPU. Note that the original StoreGPU version employs blocking data transfer operations, which hinder the ability to overlap data transfer operations and kernel executions.

For all data sizes, the ratio of time spent on memory allocations on both the GPU and the CPU is high (up to 85% for small data blocks and at least 38% for the larger data sizes). Further, the figure demonstrates that the corresponding execution pipeline (i.e., copy in, kernel, and copy out) is not balanced as the kernel execution time dominates the pipeline. Still, as the data size increases, the proportional time spent on the copy in stage becomes closer to that spent on kernel execution time, making the pipeline more balanced.

Figure 7 shows the speedup obtained when using CrystalGPU framework for a stream of 10 jobs compared to the performance of the original StoreGPU implementation as a baseline. The figure demonstrates that by exploiting buffer reuse, CrystalGPU is able to amortize memory management costs; enabling, as a result, more than 6x speedup for small data blocks (corresponding to 85% savings), and more than 1.6x for larger data sizes (corresponding to 38% savings).

Enabling the overlap feature allows for an additional speedup increase that corresponds to the portion of time spent on data transfer operations (up to 8x for small data blocks and at least 3x for larger data blocks).

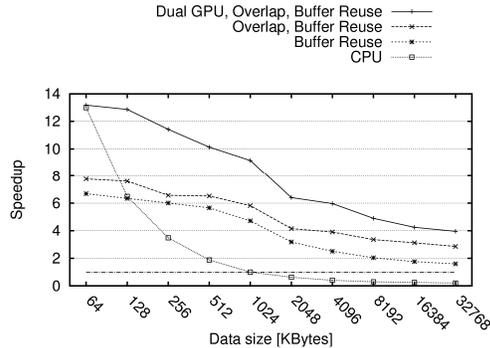

**Figure 7: Achieved speedup for SHA1 for a stream of 10 jobs. The baseline is the original StoreGPU implementation (values below 1 are slowdown). The figure also presents the performance when run on one CPU core.**

Also, using both GPUs available in the card increases the speedup further, this time by exploiting additional resources. For small data blocks, the speedup almost doubles; however, for larger blocks the speedup gains are limited by two factors. First, the ratio of time spent on data transfers increases as we increase the data size, thus placing more contention on the shared I/O channel. The second factor relates to the increased ratio of time spent on the post-processing stage, which is executed on the CPU.

We note that the absolute kernel execution time of a single SHA1 hashing job ranges from 0.5ms for the smallest data size to 10ms for the largest. In all cases, the kernel execution time is much larger than the 70μs overhead introduced by the framework, hence, rendering this overhead insignificant.

It is interesting to note that while the original StoreGPU performance lags behind the CPU performance for small data sizes (less than 1MB), enabling all Crystal GPU optimizations achieves significant speedup for all data sizes.

Figure 8 explores the achieved speedup for a stream of jobs that hashes a block of 1MB while varying the stream length. For a stream of one job, the only feature that brings speedup is buffer reuse, which is intuitive. However, increasing the stream size enables the other two features to contribute additional speedup, while buffer reuse has constant effect. On the one hand, the overlap feature is able to hide the relatively small percentage of time spent on data transfers (Figure 6), increasing the speedup by up to 1x (from 7x to 8x). On the other hand, enabling the dual GPU feature almost doubles the speedup immediately after increasing the stream size above 1.

The above experiment demonstrates that long streams are not required to benefit from most of the speed up enabled by the framework, hence making the framework beneficial even for bursty workloads.

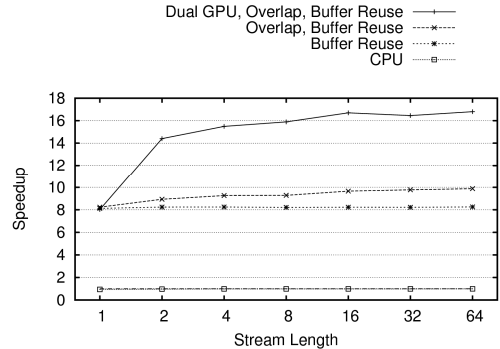

**Figure 8: Achieved speedup for SHA1 hashing for a block of 1MB while varying the stream length (values below 1 are slowdown). Note that for this data size, the CPU and the original StoreGPU implementation have the exact same performance.**

## 6 Related Work

Using GPUs for general purpose programming has recently gained considerable attention. This has led to the emergence of a number of frameworks and development environments that simplify the task of software developers to program algorithms for GPUs. Examples of such development frameworks include NVIDIA's CUDA [13], Apple's OpenCL [15] and RapidMind Development Platform [14], in addition to environments developed in academia such as BrookGPU [16] and Sh [17].

Our work is different in that it operates at a higher level than the above-mentioned development frameworks as it uses the services provided by such tools. Specifically, CrystalGPU provides a higher level abstraction that aims to transparently exploit a number of opportunities while viewing the GPU as an abstract computing device.

Building frameworks and higher level abstractions is an approach often adopted to extend the lower level mechanisms with semantics required by the higher level layers. For instance, frameworks could be built to provide priority based scheduling for computing, storage or networking resources [18], or to present a scalar computing device as a vector processor. Among the frameworks related to our work, Thompson et. al. [19] develop a framework to abstract graphical processing GPUs (GPUs without general purpose computing support) as vector processing accelerators, effectively hiding the graphical processing pipeline details and enabling general purpose computation on GPUs that do not support general purpose computing.

## 7 Summary and Future Work

We have described CrystalGPU, an execution framework that enables GPGPU applications to transparently maximize the utilization of the GPU. The framework is based on three opportunities. First, reusing GPU memory buffers to reduce the cost of allocation/deallocation of short lived buffers. Second, overlapping communication and computation to hide the communication overhead. Third, harnessing the computational power of multi-GPU systems.

Through a use-case application, namely StoreGPU, we demonstrate that CrystalGPU is able to efficiently facilitate the above-mentioned opportunities enabling an up to 16x speedup.

The results CrystalGPU achieved encourage us extend CrystalGPU framework with a more flexible scheduling mechanism. Current CrystalGPU version implements simple FIFO scheduling policy between jobs. However, some applications may require prioritizing the processing of specific jobs or enforce processing deadlines. For instance, in MPEG video encoding, the application may prefer to process the "I" frames (the encoding base frames) at higher priority. Similarly, a deep packet inspection application may require prioritized processing of packets that belong to certain classes of service. To this end, we aim to extend CrystalGPU's interface to enable passing application specific information that allows the application to customize job scheduling, hence enhancing application's performance.